\documentstyle[12pt]{article}
\begin{document}
\centerline{{\large {\bf An Approach to Master Symmetries}}} 
\centerline{{\large {\bf of Lattice Equations}}} 

\vskip 2mm
\centerline{{Benno Fuchssteiner$^\dagger$\footnote{Email: 
benno@uni-paderborn.de} 
and Wen-Xiu Ma$^\ddagger$\footnote{Email: wenxiuma@uni-paderborn.de}}} 
\centerline{$^\dagger${\small  
Universit\"at-GH Paderborn, D-33098 Paderborn, Germany}}
\centerline{$^\ddagger${\small Dept. of Math., University of Manchester 
Institute of Science and Technology,}} 
\centerline{{\small Manchester M60 1QD, UK}}   

\newcommand{\eqnsection}{
   \renewcommand{\theequation}{\thesection.\arabic{equation}}
   \makeatletter
   \csname $addtoreset\endcsname
   \makeatother}
\eqnsection

\newtheorem{thm}{Theorem}[section]
\newtheorem{ex}{Example}[section]
\newtheorem{defi}{Definition}[section] 
\newcommand{\R}{\mbox{\rm I \hspace{-0.9em} R}}   
\def\be{\begin{equation}}
\def\ee{\end{equation}}
\def\ba{\begin{array}}
\def\ea{\end{array}}
\def\bea{\begin{eqnarray}}
\def\eea{\end{eqnarray}}
\def\la {\lambda}
\def \part {\partial}
\def \al {\alpha}
\def \de {\delta}

\setlength{\baselineskip}{15pt}

\begin{abstract}
An approach to master symmetries of lattice equations 
is proposed by the use of discrete zero curvature equation. 
Its key is to generate non-isospectral flows from the discrete
spectral problem associated with a given lattice equation.
A Volterra-type lattice hierarchy and the Toda lattice hierarchy 
are analyzed as two illustrative 
examples.
\end{abstract}

\section{Introduction}
\setcounter{equation}{0}

Symmetries are one of important 
aspects of soliton theory. 
When any integrable character hasn't been found   
for a given equation, among the most 
efficient ways is to consider its symmetries. 
It is through symmetries that Russian scientists 
et al. classified many integrable equations including
lattice equations \cite{MikhailovSS} \cite{LeviY}.
They gave some specific description 
for the integrability of nonlinear equations in terms of symmetries,
and showed that if an equation possesses higher  
differential-difference degree symmetries, 
then it is subject to certain conditions, for example, the degree of 
its nonlinearity mustn't be too large, compared with its 
differential-difference degree. Usually an integrable equation
in soliton theory
is referred as to an equation possessing infinitely many symmetries 
\cite{FuchssteinerF} \cite{Fokas}. 
Moreover these symmetries form beautiful algebraic structures
\cite{FuchssteinerF} \cite{Fokas}.       

The appearance of master symmetries \cite{Fuchssteiner1983}
gives a common character for integrable differential equations 
both in $1+1$ dimensions and in $1+2$ 
dimensions, for example, the KdV equation and the KP equation. 
The resulting symmetries are sometimes
called $\tau$-symmetries \cite{Ma1990} and 
constitute centreless Virasoro algebras together
with time-independent symmetries \cite{ChenL}.  Moreover this kind of
$\tau$-symmetries may be generated by use of 
Lax equations \cite{Ma1992b} or zero curvature equations 
\cite{Ma1993b}.
In the case of lattice equations, there also exist
some similar results. For instance, a lot of lattice equations have 
$\tau$-symmetries 
and centreless Virasoro symmetry algebras
\cite{OevelFZ}.                                         
So far, however, there has not been a systematic theory to 
construct this kind of $\tau$-symmetries for lattice equations.

The purpose of this paper   
is to provide a procedure to generate those master symmetries 
for a given lattice hierarchy. The discrete zero
 curvature equation is our basic tool to give such a 
 procedure.  
A Volterra-type lattice hierarchy and the Toda lattice hierarchy 
are chosen and
analyzed as two illustrative examples, which have one 
dependent variable and two dependent variables, respectively.

The paper begins by discussing 
discrete zero curvature equations.
It will then go on to 
establish an approach to master symmetries by using
discrete zero curvature equations.
The fourth section will give rise to applications of our approach 
to two concrete examples of lattice hierarchies.
Finally, the fifth section 
provides some concluding remarks. 

\section{{Discrete zero curvature equations}}
\setcounter{equation}{0}

Let $u(n,t)$ be a function defined over $Z\times \R$,
$E$ be a shift operator: $(Eu)(n)=u(n+1)$, and $K^{(m)}=E^mK, \
m\in Z, \ K$ being a vector function.
Consider the discrete spectral problem
\be \left \{ \ba {l} E\phi =U\phi=U(u, \la )\phi ,
 \vspace{2mm} \\ \phi_t=V\phi=V(u,\la )\phi, \ea \right.
\label{sp}\ee
where $U,V$ are the same order square matrix operators and $\la $ is 
a spectral parameter.
Its integrability condition is the following discrete zero curvature
equation 
\bea  U_t&=&(EV)U-UV =((E-1)V)U-UV+VU\nonumber
\\ &=& ((E-1)V)U-[U,V]=(\Delta _+V)U-[U,V].
\eea
Recall that the continuous zero curvature equation reads as
\[ U_t=V_x-[U,V].\]
Therefore we see that there is 
a slight difference between two zero curvature equations.
The Gateaux derivative of $X(u)$ at a 
direction $S(u)$ is defined by 
\be X'[S]=\left.\frac d {d\varepsilon}\right|_{\varepsilon =0}
X(u+\varepsilon S).\ee
Throughout the paper, we assume that the spectral operator $U$ has  the 
 injective property, that is, if $U'[K]=0$,
then $K=0$. Therefore
the Gateaux derivative $U'$ is an injective linear map.
For example, in the case of Toda lattices, the spectral operator $U$  
reads as \cite{Tu}
\be 
 U=\left ( \ba {cc}0&1\vspace{2mm}\\-v&\la -p \ea \right), \ u=
\left ( \ba {c}p\vspace{2mm}\\v\ea \right), \label{Todasp}\ee
and thus we have
 \[ U'[K]=
\left (\ba {cc}0&0\vspace{2mm}\\-K_2&-K_1 \ea \right)=0 \Rightarrow K
=\left ( \ba {c}K_1\vspace{2mm}\\K_2\ea \right)=0,
 \]
 which means that the spectral operator of the Toda lattice hierarchy has the
 injective property.
We also need another property of the spectral operator 
\be {\rm if}\  (EV)U-UV=U'[K]\ {\rm and}\  V|_{u=0}=0, \ {\rm then}\  V=0,
\label{unip}\ee
which is called the uniqueness property.
Further we obtain $K=0$ from (\ref{unip}) by
the injective property. It may be shown that (\ref{Todasp}) 
share such a property, when $V$ is Laurent-polynomial dependent on $\la $. 

Note that  $U_t=U'[u_t]+\la _tU_\la =U'[u_t]+f(\la )U_\la$
when $\la _t=f(\la ),$ where $U_\la =\frac {\part U}{\part \la }$. If 
$(V,K,f)$ satisfies a so-called key discrete zero curvature equation
\be (EV)U-UV=U'[K]+fU_\la ,
 \label{DZCe}\ee 
 then we can say that when $\la _t=f(\la )$,
\[\ba {ccc} u_t=K(u)&\Longleftrightarrow  &U_t=(EV)U-UV. \\
{ {{\rm a\  discrete\ evolution\  equation}}}& &{\rm {{ 
a\  discrete\  zero\  curvature\   equation}}}
 \ea \]
 This result may be proved as follows.

\noindent {\bf Proof:} ($\Rightarrow$) \[ 
U_t=U'[u_t]+\la _tU_\la 
=U'[K]+fU_\la 
=(EV)U-UV. \]
($\Leftarrow$) \[\ba {l} U'[K]+fU_\la =(EV)U-UV=U_t =U'[u_t]+\la _tU_\la 
\vspace{2mm}\\ =U'[u_t]+f U_\la  \Rightarrow 
  U'[u_t-K]=0  \Rightarrow  u_t=K,\ea \]
  where the linearity of $U'$ 
  and the injective property of $U$ are used in the last two steps.
  $\vrule width 1mm height 3mm depth 0.4mm$        

\begin{defi} 
A matrix operator $V$ is called a Lax operator corresponding 
to the spectral operator
$U$ with a spectral evolution law $\la _t =f(\la )$
if a key discrete zero curvature equation (\ref{DZCe}) holds. 
Moreover $V$ is called an isospectral Lax operator if $f=0$ or 
a nonisospectral Lax operator if $f\ne 0$.
\end{defi}

The equation (\ref{DZCe}) 
exposes an essential relation between a discrete equation and 
its discrete zero curvature representation. It 
will play an important role in the context of our construction
of master symmetries.

\section{{ \bf An approach to master symmetries}}
\setcounter{equation}{0}

What are master symmetries? For a given evolution equation $u_t=K(u)$,
where $K(u)$ does not depend 
explicitly on $t$, the definition of master symmetries 
is the following \cite{Fuchssteiner1983}.
\begin{defi} 
A vector field $\rho (u)$ is called a master symmetry of $u_t=K(u)$
if 
$ [K,[K,\rho ]]=0$, where the commutator of two vector fields is defined by 
\be [K,\rho]=K'[\rho ]-\rho'[K].\ee  
\end{defi}

If $\rho (u)$ is a master symmetry of $u_t=K(u)$, then 
the vector field 
$\tau =t[K,\rho]+\rho  $,  depending explicitly on the time $t$,
is a symmetry of $u_t=K(u)$, 
namely, to satisfy the linearized equation of $u_t=K(u)$:
\be \frac {d \tau }{d t}=K'[\tau ],\ i.e.\  \frac {\part \tau }{\part t}=
[K,\rho].\ee

Main Observation: 

\centerline{\it Nonisospectral  flows  with  $\la _t=\la ^{n+1}$
$\Rightarrow$  master 
symmetries.}
\noindent So for $(K,V,f)$ and $(S,W,g)$,
we introduce 
a new product 
\be
\lbrack\!\lbrack V, W\rbrack\!\rbrack =V'[S]-W'[K]+[V,W]+gV_\la 
-fW_\la ,\label{prodVW}
\ee
in order to discuss master symmetries.
An algebraic structure for the key discrete zero curvature 
equation is shown in the following theorem.
\begin{thm} \label{AlgS} If two discrete zero curvature equations
\bea && (EV)U-UV=U'[K]+fU_\la ,\label{eq1}\\ && 
(EW)U-UW=U'[S]+gU_\la ,\label{eq2}
\eea
hold, then we have the third discrete zero curvature equation
\be (E\lbrack\!\lbrack V,W\rbrack\!\rbrack)U-U\lbrack\!\lbrack V,W
\rbrack\!\rbrack=U'[T]+\lbrack\!\lbrack f,g\rbrack\!\rbrack U_\la ,
\label{prodp}\ee
where $T=[K,S]$ and 
 $ \lbrack\!\lbrack f, g \rbrack\!\rbrack$ 
 is defined by 
 \be \lbrack\!\lbrack f, g \rbrack\!\rbrack (\la )
=f'(\la )g(\la )-f(\la )g'(\la ).\ee
\end{thm}
\noindent {\bf Proof:} 
The proof is an application of equalities (\ref{eq1}),  
(\ref{eq2}) and   
\[ (U'[K])'[S]-(U'[S])'[K]=U'[T],\ T=[K,S],\] 
which is a similar result to one in the continuous case
in \cite{Ma1992} and may be immediately
checked.
We first observe that 
\[ ({\rm Equ.}\  (\ref{eq1}))'[S]-
({\rm Equ.}\  (\ref{eq2}))'[K]  
+g({\rm Equ.}\  (\ref{eq1}))_{\la }-f({\rm Equ.}
\  (\ref{eq2}))_{\la },\]
and then a direct calculation may give the equality (\ref{prodp}).
$\vrule width 1mm height 3mm depth 0.4mm$  

This theorem shows that
a product equation $u_t=[K,S]$  
has 
a discrete zero curvature representation
\[ U_t=(E   \lbrack\!\lbrack V, W \rbrack\!\rbrack  )U-
U\lbrack\!\lbrack V,W\rbrack\!\rbrack
\ \ {\rm with}\ \ \la _t
=\lbrack\!\lbrack f,g\rbrack\!\rbrack ,\]
 when two evolution equations $u_t=K(u)$ and $u_t=S(u)$ have
 the discrete zero curvature representations:
\be U_t=(EV)U-UV\ \ {\rm with}\ \ \la _t =f(\la ),
 \ U_t=(EW)U-UW\ \ {\rm with}\ \ \la _t =g(\la ),\nonumber
 \ee
 receptively.
According to the above theorem, we can also easily find that 
if an equation $u_t=K(u)$ is isospectral ($f=0$), then 
the product equation $u_t=[K,S]$ for any vector field $S(u)$
is still isospectral
because we have $\lbrack\!\lbrack f,g\rbrack\!\rbrack =
\lbrack\!\lbrack 0,g\rbrack\!\rbrack = 0$, where $g$ is the evolution law
corresponding $u_t=S(u)$
(see \cite{Ma1993} for the continuous case).

Let us now assume that we already have a hierarchy of isospectral 
equations of the form
\be u_{t}
=K_k=\Phi ^kK_0, \ k\ge 0,  \label{dlh}
\ee
associated with a discrete spectral problem 
\be E\phi =U\phi ,\ \phi =(\phi_1,\cdots, 
\phi_r)^T.\label{spx}\ee
Usually a discrete  spectral  problem $E\phi =U\phi$ yields 
a hereditary operator  $\Phi$ (see \cite{FokasA} for instance), 
i.e. a square matrix operator
to satisfy
\[\Phi^2[K,S]+\Phi [\Phi K, \Phi S]-\{\Phi[K,\Phi S]+\Phi 
[\Phi K, S]\}=0 \]
for any vector fields $K,S$.

In order to generate nonisospectral flows, we further
introduce an operator equation of $\Omega(X)$:
\be
(E\Omega (X))U-U\Omega (X)=U'[\Phi X]-\lambda U'[X]. \label{cq}
  \ee
We call it a characteristic operator equation of $E\phi =U\phi$.

\begin{thm} \label{VWk}
Let two matrices $V_0, W_0$
and two vector fields $K_0,\rho_0$ 
satisfy 
\bea &&(EV_0)U-UV_0=U'[K_0],\label{V0} \\
 && (EW_0)U-UW_0=U'[\rho _0]+\la U_\la ,\label{W0} \eea  
 and $\Omega (X)$ be a solution to (\ref{cq}).
If we define a hierarchy of new vector fields and two hierarchies of 
square matrix operators as follows
\bea && \rho_l=\Phi ^l\rho_0,\ l\ge1,
  \label{nonidlh} \\
&& V_k=
 \lambda ^k V_0 + \sum _{i=1}^k
\lambda ^{k-i}\Omega (K_{i-1})\ 
,\ k\ge 1, \label{Vk}
\\ &&W_l=
 \lambda ^l W_0 + \sum _{j=1}^l
\lambda ^{l-j}\Omega (\rho _{j-1})\ 
,\  l\ge 1,\label{Wk}
\eea
then the square matrix operators $V_k,W_l,\, k,l\ge 0$, satisfy
\bea && (EV_k)U-UV_k=U'[K_k],\ k\ge 0,
\label{VkU}\\ && (EW_l)U-UW_l=U'[\rho_l]+\la ^{l+1}U_\la 
,\ l\ge 0,\label{WkU} \eea
respectively. Therefore  $u_t=K_k$
and $u_t=\rho_l$
possess the isospectral ($\lambda _t=0$) and nonisospectral ($\la 
_t=\la ^{l+1}$) discrete zero curvature representations
\[ U_t=(EV_k)U-UV_k,\  U_t=(EW_l)U-UW_l,\]  respectively.
\end{thm}
  
\noindent {\bf Proof:} 
We prove two equalities (\ref{VkU}) and (\ref{WkU}).
We can compute that
\bea 
&&(EV_k)U-UV_k \nonumber \\     
&=& \la ^k[(EV_0)U-UV_0]+\sum_{i=1}^k\la ^{k-i}[E\Omega (K_{i-1})U- 
U\Omega (K_{i-1})]\nonumber \\     
&=& \la ^k U'[K_0]+\sum_{i=1}^k\la ^{k-i}\{U'[\Phi K_{i-1}]
-\la U'[K_{i-1}]\}\nonumber \\     
&=& \la ^kU'[K_0]+\sum_{i=1}^k\la ^{k-i}\{ U'[K_i]-\la U'[K_{i-1}]\}
\nonumber \\
& = & U'[K_k],\ k\ge 1;\nonumber \vspace{2mm}\\     
& & (EW_l)U-UW_l \nonumber \\     
&=& \la ^l[(EW_0)U-UW_0]+\sum_{j=1}^l\la ^{l-j}[E\Omega (\rho _{j-1})U- 
U\Omega (\rho_{j-1})]\nonumber \\     
&=& \la ^l\{ U'[\rho _0]+\la U_{\la }\} 
+\sum_{j=1}^l\la ^{l-j}\{ U'[\Phi \rho _{j-1}]-\la U'[\rho _{j-1}]\}
\nonumber \\
&=& \la ^l\{ U'[\rho _0]+\la U_{\la }\} 
+\sum_{j=1}^l\la ^{l-j}\{ U'[\rho _j]-\la U'[\rho _{j-1}]\}
\nonumber \\
&= & U'[\rho _l]+\la ^{l+1}U_{\la },\ l\ge 1.\nonumber    
\eea
Note that we have used the characteristic operator equation (\ref{cq}). 
The rest is obviously and the proof is therefore finished.
$\vrule width 1mm height 3mm depth 0.4mm$  

This theorem gives rise to the structure of Lax operators associated with
the isospectral and nonisospectral hierarchies.
In fact, 
the theorem provides us with a method to 
construct an isospectral hierarchy and a nonisospectral hierarchy associated
with a discrete spectral problem (\ref{spx})  
by solving two initial key discrete zero 
curvature equations (\ref{V0}) and (\ref{W0}) and by solving 
a characteristic operator  equation
 (\ref{cq}), if a hereditary operator is known. However, the hereditary 
 operator $\Phi$ and the isospectral hierarchy (\ref{dlh})
 are often determined
 from the spectral problem (\ref{spx}) simultaneously. Therefore 
 we obtain just a new nonisospectral hierarchy (\ref{nonidlh}). 
A operator solution to  (\ref{cq}) 
may be generated by changing the element $K_k$ (or $G_k$) in the 
following equality
\be \Omega (K_k) =V_{k+1}-\la V_k,
\label{omegas}  \ee
which may be checked through (\ref{Vk}).
The Lax operator 
matrices $V_{k+1}$ and $ V_k$ are known, when the isospectral 
hierarchy has 
already been given.
Therefore the whole process of construction of nonisospectral hierarchies
becomes an easy task:
finding $\rho _0,W_0$ to satisfy (\ref{W0}) 
and computing $V_{k+1}-\la V_k$.

The nonisospectral hierarchy (\ref{nonidlh}) is exactly the required
master symmetries. The reasons
are that the product equations of the isospectral hierarchy and 
the nonisospectral hierarchy are still isospectral by Theorem \ref{AlgS},
and the isospectral 
equations often commute with each other.
Therefore it is because there exists a nonisospectral hierarchy that
there exist master symmetries for lattice equations derived from 
a given spectral problem.
In the next section, we will in detail 
explain our construction process by two 
concrete examples and establish their
corresponding centreless Virasoro symmetry algebras.

\section{{ \bf Application to lattice hierarchies}}
\setcounter{equation}{0}

We explain by two lattice hierarchies how to apply 
the method in the last section to construct master 
symmetries.

\noindent {\bf Example 4.1.} {\it A Volterra-type lattice hierarchy.}
Let us first consider the discrete spectral problem \cite{ZhangTOF}:
\be E\phi =U\phi, \ U=\left ( \ba {cc} 1&u\vspace {1mm}
\\ \la ^{-1}&0\ea \right ), \ \phi = \left ( \ba {c} \phi _1\vspace{1mm}
\\\phi_2\ea \right ).\label{sp1}\ee
The corresponding isospectral lattice hierarchy:
\be 
 u_{t}=K_k=\Phi ^{k}K_0=u(a_{k+1}^{(1)}-a_{k+1}^{(-1)}),\ 
K_0=u(u^{(-1)}-u^{(1)}),\ k\ge0, \label{dlh1}\ee
where the hereditary operator $\Phi$ is defined by
\[
\Phi=u(1+E^{-1})(-u^{(1)}E^2+u)(E-1)^{-1}u^{-1}. \nonumber  \]
The associated Lax operators are as follows
\be V_{k}= (\la ^{k+1}V)_{\ge 1}+
\left ( \ba {cc} a_{k+1}&0\vspace{2mm}
\\ c_{k+1} & a_{k+1}^{(-1)}\ea \right ), \ k\ge 0,\ee
where $(P)_{\ge i}$ denotes
the selection of the terms with degrees of $\la $ no less than $i$. 
The matrix $V=\sum_{i\ge 0}V_{(i)}\la ^{-i}=
\sum_{i\ge 0}\left ( \ba {cc} a_{i}&b_{i}
\\ c_i &-a_i\ea \right )\la ^{-i}$ solves the stationary ($U_t=0$)
zero curvature equation $(EV)U-UV=0$, and we choose 
$a_0=\frac12 , b_0=u, c_0=0, \, a_1=-u,b_1=-(u^2+uu^{(-1)}),c_1=1$ 
and require that $a_{i+1}|_{u=0}=b_{i+1}|_{u=0}=c_{i+1}|_{u=0}=0,\,
i\ge 1$. Actually  we have 
\be \left \{ \ba {l} 
a_{i+1}^{(1)}-a_{i+1}=-u^{(1)}c^{(2)}_{i+1}+uc_{i+1}, 
\ i\ge 1,\vspace{2mm}\\
b_{i+1}=uc_{i+2}^{(1)},\ i\ge 1\vspace{2mm}\\
c_{i+1}=a_i+a_i^{(-1)},\ i\ge 1. \ea \right. \label{recursionr1}\ee
In particular, we can obtain
 \[ V_0=\left ( \ba {cc} \frac12 \la -u&\la u\vspace{2mm}
\\ 1 &-\frac1 2\la -u^{(-1)}\ea \right ).\] 

\noindent {\bf Nonisospectral Hierarchy:}

\noindent \underline {Step 1:}
To solve the nonisospectral 
($\la _t=\la $) initial key 
discrete zero curvature equation (\ref{W0}) yields a pair of  solutions:
\be \rho_0=u,\ W_0=\left ( \ba {cc} \frac12 & 0\vspace{2mm}
\\ 0 &-\frac1 2\ea \right ).
 \ee
 
\noindent \underline {Step 2:}  
We compute that 
\[ \ba {l} \ \  V_{k+1}-\la V_k \vspace{2mm}\\
= (\la ^{k+2}V)_{\ge 1}+
\left ( \ba {cc} a_{k+2}&0\vspace{2mm}
\\ c_{k+2} & a_{k+2}^{(-1)}\ea \right )
-\la (\la ^{k+1}V)_{\ge 1}-\la 
\left ( \ba {cc} a_{k+1}&0\vspace{2mm}
\\ c_{k+1} & a_{k+1}^{(-1)}\ea \right )\vspace{2mm} \\
=\left ( \ba {cc} a_{k+2}&\la b_{k+1}\vspace{2mm}
\\ c_{k+2} & -\la (a_{k+1}+a_{k+1}^{(-1)})+
a_{k+2}^{(-1)}\ea \right ).\nonumber \ea \]
On the other hand, by (\ref{dlh1}) and (\ref{recursionr1}), we have 
\[ \left \{ \ba {l} 
a_{k+1}=(E-E^{-1})^{-1}u^{-1} X_{k}, \vspace{2mm} \\    
a_{k+2}=(E-E^{-1})^{-1}u^{-1} \Phi X_{k},\vspace{2mm} \\ 
c_{k+2}=a_{k+1}+a_{k+1}^{(-1)}=(1+E^{-1})a_{k+1},\vspace{2mm}\\
b_{k+1}=uc_{k+2}^{(1)}=u(E+1)a_{k+1}.\ea \right.
\]
Now by interchanging the element  
$X_k$ into $X$ in the quantity $V_{k+1}-\la V_k$, we obtain 
a solution to the corresponding characteristic operator equation:
\be \Omega (X)=\left ( \ba {cc} \Omega _{11}(X)&\Omega 
_{12}(X)\vspace{2mm}\\
\Omega _{21}(X)&\Omega _{22}(X)\ea \right ),
\ee 
where $\Omega_{ij}(X)$, $i,j=1,2$, are given by 
\bea &&
\Omega _{11}(X)=
(E-1)^{-1}(-u^{(1)}E^2+u)(E-1)^{-1}u^{-1}X\nonumber \\ &&
 \Omega _{12}(X)= \la uE(E-1)^{-1}u^{-1}X\nonumber \\ &&
\Omega _{21}(X)= (E-1)^{-1}u^{-1}X 
\nonumber \\ &&
\Omega _{22}(X)=[-\la +E^{-1}(E-1)^{-1}
(-u^{(1)}E^2+u)](E-1)^{-1}u^{-1}X\nonumber .
\eea 
Therefore we obtain a hierarchy of nonisospectral discrete
equations $ u_t=\rho_l=\Phi^l\rho_0,\ l\ge0,$ by Theorem \ref{VWk}.

\noindent {\bf Symmetry Algebra:}

\noindent \underline {Step 3:}
We make the following computation at $u=0$:
\bea && K_k|_{u=0}=0,\ \rho _l|_{u=0}=
\Phi ^l\rho_0|_{u=0}=0
,\ k,l\ge0,\nonumber\\
&&
V_k|_{u=0}=\la ^k\left (\ba {cc} \frac12 \la & 0\vspace{2mm}\\ 1& 
-\frac 12 \la \ea \right),\ k\ge0,\nonumber\\ && 
W_l|_{u=0}=\la ^l\left (\ba {cc} \frac12  & 0 \vspace{2mm} \\ 
0 & -\frac 12  \ea \right )+
(1-\delta_{l0}) \la ^{l-1} 
\left (\ba {cc} 0 & 0 \vspace{2mm}  \\   
{[n]}& -\la [n]  \ea \right )
,\ l\ge0,\nonumber
\eea 
where $\delta_{l0}$ represents the Kronecker symbol,
and $[n]$ denotes a multiplication operator
$[n]: u\mapsto [n]u$, $([n]u)(n)=nu(n),$
involved in the construction of nonisospetral hierarchies. 
While computing $W_l|_{u=0}$, we need to note that
$\Omega(\rho _0)|_{u=0}\ne 0,$ but 
$\Omega(\rho _l)|_{u=0}= 0,\ l\ge1.$ 
Now we may find by the definition (\ref{prodVW}) of the product of two
Lax operators that
\be \left \{ \ba {l} 
\lbrack\!\lbrack V_k,V_l\rbrack\!\rbrack|_{u=0}=0,\ k,l\ge 0,\vspace{2mm}\\
\lbrack\!\lbrack V_k,W_l\rbrack\!\rbrack|_{u=0}=(k+1)V_{k+l}|_{u=0},
\ k,l\ge 0,\vspace{2mm}
\\
\lbrack\!\lbrack W_k,W_l\rbrack\!\rbrack|_{u=0}=(k-l)W_{k+l}|_{u=0},
\ k,l\ge 0.\ea \right.\nonumber\ee

\noindent \underline {Step 4:} It is easy to prove that 
\[ \lbrack\!\lbrack V_k,V_l\rbrack\!\rbrack, 
\lbrack\!\lbrack V_k,W_l\rbrack\!\rbrack-(k+1)V_{k+l}, 
\lbrack\!\lbrack W_k,W_l\rbrack\!\rbrack-(k-l)W_{k+l},\ k,l\ge0,\]
are all isospectral ($\la _t=0$) Lax operators.
For example, the spectral evolution laws of the Lax operators
of the third kind
are 
\[ \lbrack\!\lbrack \la ^{k+1},\la ^{l+1}\rbrack\!\rbrack-(k-l)\la ^{k+l+1}
=(k+1)\la ^k\la ^{l+1}-(l+1)\la ^{k+1}\la ^l-(k-l)\la ^{k+l+1}=0.\]
Then by the uniqueness property of the spectral problem (\ref{sp1}),
we obtain a Lax operator algebra  
\be \left \{ \ba {l}
\lbrack\!\lbrack V_k,V_l\rbrack\!\rbrack =0,\ k,l\ge 0,\vspace{2mm}\\
\lbrack\!\lbrack V_k,W_l\rbrack\!\rbrack =(k+1)V_{k+l},
\ k,l\ge 0,\vspace{2mm}\\
\lbrack\!\lbrack W_k,W_l\rbrack\!\rbrack=(k-l)W_{k+l},
\ k,l\ge 0.
\ea\right.\label{laxa1}\ee
Further, due to the injective property of $U'$,
we finally obtain a vector field algebra of the isospectral hierarchy 
and the nonisospectral hierarchy
\be \left \{ \ba {l}
\left[ K_k,K_l \right] =0,\ k,l\ge 0,\vspace{2mm}\\
\left[ K_k,\rho _l \right] =(k+1)K_{k+l},
\ k,l\ge 0,\vspace{2mm}\\
\left[ \rho _k,\rho _l\right]=(k-l)\rho _{k+l},
\ k,l\ge 0.
\ea \right. \label{sa1} \ee
This implies that $\rho _l,\ l\ge 0$, are all master symmetries of 
each equation $u_t=K_{k_0}$ 
of the isospectral hierarchy, and the symmetries 
\be K_k,\,k\ge 0,\   
\tau _l^{(k_0)}=t[K_{k_0},\rho_l]+\rho_l=(k_0+1)tK_{k_0+l}+\rho_l,
\ l\ge0, \ee
constitute  a symmetry algebra of Virasoro type possessing
the same commutator relations as (\ref{sa1}).

\noindent {\bf Example 4.2:} {\it The Toda lattice hierarchy.}
Let us second consider the discrete spectral problem \cite{Tu}:
\be E\phi =U\phi, \ U=\left ( \ba {cc} 0&1\vspace {1mm}
\\ -v&\la -p\ea \right ), \ u=\left ( \ba {c} p\vspace{1mm}
\\ v\ea \right ),\ 
\phi = \left ( \ba {c} \phi _1\vspace{1mm}
\\\phi_2\ea \right ),\label{sp2}\ee
which is equivalent to $(E+vE^{-1}+p)\psi =\la \psi . $
The corresponding isospectral ($\la _t=0$)
integrable Toda lattice hierarchy reads as 
\be 
 u_{t}=K_k=\Phi ^{k}K_0
,\ K_0=\left ( \ba {c} v-v^{(1)}\vspace{1mm}
\\ v(p-p^{(-1)})\ea \right )  ,\ k\ge 0,
\ee 
where the hereditary operator $\Phi $ is given by 
\be 
\Phi=\left ( \ba {cc} p&(v^{(1)}E^2-v)(E-1)^{-1}v^{-1}\vspace{1mm}
\\ v(E^{-1}+1)& \ \ v(pE-p^{(-1)})(E-1)^{-1}v^{-1}\ea \right ).\nonumber\ee
The first nonlinear discrete equation is exactly the Toda lattice \cite{Toda}
\be \left \{\ba {l} p_t(n)=v(n)-v(n+1),
\vspace{2mm}\\ v_t(n)=v(n)(p(n)-p(n-1)),\ea \right. \ee
up to a transform of dependent variables.
The corresponding Lax operators read as 
\be V_{k}=(\la ^{k+1}V)_{\ge 0}+
\left ( \ba {cc} b_{k+2}&0\vspace{1mm}
\\ 0&0\ea \right ), \ k\ge0,
 \ee
Here $V=\sum_{i\ge 0}V_{(i)}\la ^i=
\sum_{i\ge0}\left ( \ba {cc} a_i&b_i\vspace{1mm}
\\ c_i& -a_i\ea \right )\la ^{-i}$ solves the stationary 
discrete zero curvature equation $(EV)U-UV=0,$ and we choose 
$ a_0=\frac12 ,\ b_0=0,\ c_0=0,\ a_1=0, \ b_1=-1,\ c_1=-v$
and require that 
$ a_{i+1}|_{u=0}=b_{i+1}|_{u=0}=c_{i+1}|{u=0}=0,\ i\ge 1.$
More precisely, we have 
\be \left \{ \ba {l} a_{i+1}^{(1)}-a_{i+1}=p(a_i^{(1)}-a_i)
+(vb_i-v^{(1)}b_i^{(2)}),\ i\ge 1,
\vspace{2mm}\\
b_{i+1}^{(1)}=pb_i^{(1)}-(a_i^{(1)}+a_i),\ i\ge 1,\vspace{2mm} \\
c_{i+1}=-vb_{i+1}^{(1)},\ i\ge 1. \ea \right. \ee
Now it is easy to find an isospectral Lax operator
\[ V_0=\left ( \ba {cc} \frac12 \la -p^{(-1)}&\ \ -1\vspace{2mm}
\\ v &\ \ -\frac1 2\la \ea \right ) .\]

\noindent {{\bf Nonisospectral Hierarchy:}}
 
\noindent \underline {Step 1:} To solve
find the nonisospectral ($\la _t=\la $)
initial key discrete zero curvature
  equation (\ref{VWk})
 leads to a pair of solutions
\[ \rho_0=\left ( \ba {c} p\vspace{2mm}
\\ 2v\ea \right ),\ W_0=\left ( \ba {cc} [n]-1 & 0\vspace{2mm}
\\ 0 &[n]\ea \right ),
 \] 
$[n]$ still being a multiplication operator
$[n]: u\mapsto [n]u$, $([n]u)(n)=nu(n).$

\noindent \underline {Step 2:}
To compute $V_{k+1}-\la V_k$
leads to a solution to the corresponding characteristic operator 
equation:
\[ \Omega (X)=\left ( \ba {cc} \Omega _{11}(X)&\Omega _{12}(X)\vspace{2mm}\\
\Omega _{21}(X)&\Omega _{22}(X)\ea \right ),\ X=\left ( \ba {c}
X_1 \vspace{2mm}\\X_2\ea \right ),
\] 
\noindent where $\Omega_{ij}(X)$, $i,j=1,2$, are given by 
\bea &&
\Omega _{11}=
E^{-1}(E-1)^{-1}X_1+(p^{(-1)}-\la )(E-1)^{-1}v^{-1}X_2,\nonumber \\ &&
 \Omega _{12}= (E-1)^{-1}v^{-1}X_2,\nonumber \\ &&
\Omega _{21}= vE(E-1)^{-1}v^{-1}X_2, 
\nonumber \\ &&
\Omega _{22}= (E-1)^{-1}X_1 \nonumber .
\eea 
At this stage, we obtain a hierarchy of nonisospectral discrete equations 
$u_t=\rho_l=\Phi ^l\rho_0,\ l\ge0,$ by Theorem \ref{VWk}.

\noindent {{\bf Symmetry Algebra:}} 

\noindent \underline {Step 3:}
We make the following computation at $u=0$:
\bea && K_k|_{u=0}=0,\ \rho _l|_{u=0}=
\Phi ^l\rho_0|_{u=0}=0,\ k,l\ge0,\nonumber\\
&&
V_k|_{u=0}=\la ^k\left (\ba {cc} \frac12 \la & -1\vspace{2mm}\\ 0& 
-\frac 12 \la \ea \right),\ k\ge0,\nonumber\\&& 
W_l|_{u=0}=\la ^l\left (\ba {cc} [n]-1  & 0\vspace{2mm}\\ 0& 
[n]  \ea \right )+
(1-\delta_{l0})\left (\ba {cc} -2\la [n]  & 2[n]\vspace{2mm}\\ 0& 
0  \ea \right )
,\ l\ge 0.\nonumber
\eea 
Now we may similarly find by the product definition (\ref{prodVW})
of the product of two Lax operators that 
\[ \left \{\ba {l}
\lbrack\!\lbrack V_k,V_l\rbrack\!\rbrack|_{u=0}=0,\ k,l\ge 0,\vspace{2mm}\\
\lbrack\!\lbrack V_k,W_l\rbrack\!\rbrack|_{u=0}=(k+1)V_{k+l}|_{u=0},
\ k,l\ge 0,\vspace{2mm}\\
\lbrack\!\lbrack W_k,W_l\rbrack\!\rbrack|_{u=0}=(k-l)W_{k+l}|_{u=0},
\ k,l\ge 0.\ea \right.\]

\noindent \underline {Step 4:} 
Because $\lbrack\!\lbrack V_k,V_l\rbrack\!\rbrack, \ 
\lbrack\!\lbrack V_k,W_l\rbrack\!\rbrack-(k+1)V_{k+l},  \ 
\lbrack\!\lbrack W_k,W_l\rbrack\!\rbrack-(k-l)W_{k+l},\, k,l\ge 0,$
are still isospectral ($\la _t=0$) Lax operators,   
a Lax operator algebra may be similarly  obtained 
by using the uniqueness property:
\be \left \{ \ba {l}
\lbrack\!\lbrack V_k,V_l\rbrack\!\rbrack =0,\ k,l\ge 0,\vspace{2mm}\\
\lbrack\!\lbrack V_k,W_l\rbrack\!\rbrack =(k+1)V_{k+l},
\ k,l\ge 0,\vspace{2mm}\\
\lbrack\!\lbrack W_k,W_l\rbrack\!\rbrack=(k-l)W_{k+l},
\ k,l\ge 0.
\ea\right.\ee
Further, through the injective property of $U'$,
a vector field algebra is yielded
\be \left \{ \ba {l}
\left[ K_k,K_l \right] =0,\ k,l\ge 0,\vspace{2mm}\\
\left[ K_k,\rho _l \right] =(k+1)K_{k+l},
\ k,l\ge 0,\vspace{2mm}\\
\left[ \rho _k,\rho _l\right]=(k-l)\rho _{k+l},
\ k,l\ge 0.
\ea \right. \label{Todava}\ee
This shows that the symmetries for $u_t=K_{k_0}$: 
\be K_k,\ k\ge 0,\ 
\tau _l^{(k_0)}=t[K_{k_0},\rho_l]+\rho_l=(k_0+1)tK_{k_0+l}+\rho _l
,\ l\ge0, \ee
constitute the same centreless Virasoro algebra as 
(\ref{Todava}).

\section{Concluding remarks} 
\setcounter{equation}{0}

We introduced a simple 
procedure to construct master symmetries for 
isospectral lattice 
hierarchies associated with discrete spectral problems.
The crucial points are an algebraic structure related to  
discrete zero curvature equations and 
the structure of
Lax operators of isospectral and
nonisospectral lattice hierarchies. 
The procedure may be divided into four steps.
The first two steps yields 
the required nonisospectral lattice hierarchy, by
solving an initial key nonisospectral discrete zero curvature equation,
and by computing a difference $V_{k+1}-\la V_k$
between two isospectral Lax operators $V_{k+1}$ and $V_k$.
The second two steps yields the  required centreless Virasoro
symmetry algebra, by using an algebraic relation between two discrete
zero curvature equations, and the uniqueness property and 
the injective property of the corresponding spectral operators.
Two lattice hierarchies
are shown as illustrative
examples. 

There is 
a common Virasoro algebraic structure for symmetry algebras 
of a Volterra-type lattice hierarchy
and the Toda lattice hierarchy.
In general, we have 
\be
[K_k,\rho_l]=(k+\gamma )K_{k+l},\ 
\gamma={\it  {const.}},\quad {\it  {{ (semi-direct\ product)}}},\ee
but the other two equalities do not change:
\be \left\{ \ba {l} [K_k,K_l]=0
\vspace{2mm}\\ 
\left [\rho_k,\rho_l \right]=(k-l)\rho_{k+l}\quad\ \ \ea \right. \ba {l}
{\it (Abelian \  algebra)},\vspace{2mm}\\
{\it (centreless\ Virasoro\  algebra)}.
\ea  \ee
This 
is also a common characteristic for continuous integrable hierarchies
\cite{Ma1992jmp}. 
Interestingly, this kind of centreless Virasoro algebra itself 
may also be used to generate variable-coefficient integrable equations 
which possess higher-degree polynomial time-dependent symmetries \cite{MaBC}.

We point out that there are some other methods 
to construct master symmetries
 by directly searching for some suitable 
vector fields $\rho _0$ 
\cite{OevelFZ} \cite{CherdantsevY} \cite{FuchssteinerIW}.
However the approach above takes full advantage 
of discrete zero curvature equations,
and therefore it is more systematic and 
may be easily applied to other lattice hierarchies.

\vskip 3mm
\noindent{\bf Acknowledgments:} One of the authors (W. X. Ma) 
would like to thank the 
Alexander von Humboldt Foundation for financial support.
He is also grateful 
to Prof. R. K. Bullough and Dr. P. J. Caudrey for their warm 
hospitality at UMIST
and to Prof. W. I. Fushchych for his kind and helpful conversations.

\end{document}